\documentclass[12pt]{iopart}

\usepackage{graphicx}
\usepackage{enumerate}
\usepackage[utf8]{inputenc}
\usepackage{cancel}
\usepackage{color}
\usepackage{hyperref}
\usepackage{float}

\usepackage{cite}
\usepackage{xr}

\newcommand{\eq}{\textnormal{\scriptsize eq}}

\newcommand{\fin}{\textnormal{\scriptsize f}}

\newcommand{\refer}{\textnormal{\scriptsize ref}}
\newcommand{\CA}{\textnormal{{\scriptsize CA}}}
\newcommand{\Carnot}{\textnormal{{\scriptsize C}}}

\def\dbar{{\mathchar'26\mkern-12mu d}}

\usepackage{iopams}  
\begin{document}

\title[Building an irreversible Carnot-like heat engine]{Building an irreversible Carnot-like heat engine with an overdamped
  harmonic oscillator}

\author{Carlos A. Plata}
\address{Dipartimento di Fisica e Astronomia ``Galileo Galilei'',
  INFN, Università di Padova, Via Marzolo 8, 35131 Padova, Italy}
\author{David Guéry-Odelin}
\address{Laboratoire Collisions, Agrégats, Réactivité, IRSAMC, Université de Toulouse, CNRS, UPS, Toulouse, France}
\author{Emmanuel Trizac}
\address{Universit\'e Paris-Saclay, CNRS, LPTMS, 91405, Orsay, France.}
\author{A. Prados}
\address{Física Teórica, Universidad de Sevilla, Apartado de
  Correos 1065, E-41080 Sevilla, Spain}


\begin{abstract}
  We analyse non-equilibrium Carnot-like cycles built with a colloidal
  particle in a harmonic trap, which is immersed in a fluid that acts
  as a heat bath. Our analysis is carried out in the overdamped
  regime. The cycle comprises four branches: two isothermal
  processes and two \textit{locally} adiabatic ones. In the latter,
  both the temperature of the bath and the stiffness of the harmonic
  trap vary in time, but in such a way that the average heat
  vanishes for all times.  All branches are swept at a finite rate
  and, therefore, the corresponding processes are irreversible, not
  quasi-static. Specifically, we are interested in optimising the heat
  engine to deliver the maximum power and characterising the
  corresponding values of the physical parameters. The efficiency at
  maximum power is shown to be very close to the Curzon-Ahlborn bound
  over the whole range of the ratio of temperatures of the two
  thermal baths, pointing to the near optimality of the proposed protocol.
\end{abstract}

%
%
%
%
%

\section{Introduction}

The investigation of heat engines is a pillar of classical
thermodynamics~\cite{callen_thermodynamics_1985}. The practical
interest of the conversion of thermal energy into mechanical work
led to unravel the laws of
thermodynamics. These laws have been well formulated since the 19th
century for macroscopic systems, for which fluctuations are
negligible. In this context, the Carnot heat engine has played a major
role: the Carnot cycle comprises two isothermal and two adiabatic
branches, which are swept in a quasi-static, reversible, way. This
\textit{reversible Carnot heat engine} maximises the efficiency but
the infinite time operation entails that the delivered power
vanishes. In the adiabatic branches, the system is thermally isolated
from the bath and there is no heat exchange---moreover, reversibility
implies that there is no entropy variation either.

The extension of thermodynamic results to mesoscopic systems, where
fluctuations are of paramount importance, is not straightforward ;  stochastic
thermodynamics has been developed 
to this end~\cite{sekimoto_stochastic_2010,seifert_stochastic_2012,ciliberto_experiments_2017},
the focus of which lies on non-equilibrium dynamics. In recent years,
researchers have looked into the possibility of speeding up the
relaxation of physical systems between two given equilibrium
states\cite{martinez_engineered_2016,muratore-ginanneschi_application_2017,li_shortcuts_2017,chupeau_engineered_2018,chupeau_thermal_2018,albay_thermodynamic_2019,albay_realization_2020}.
The ``engineered swift equilibration'' (ESE) techniques that emerged, which have also been referred to 
as ``shortcut to isothermality'' are essentially equivalent and can be viewed as the counterpart in the classical realm of the ``shortcuts to
adiabaticity'' (STA) developed in quantum
systems\cite{guery-odelin_shortcuts_2019}. Both STA and ESE processes
make it possible to connect given initial and target states in a time
that is much shorter than the natural characteristic relaxation time of the
system at hand. To avoid confusion, it is perhaps worthwhile pointing that 
``adiabaticity'' in ``STA'' refers to a slow variation, not to he absence
of heat transfer, at variance with terminology to be used below.

STA and ESE techniques make it possible to build heat engines that
connect equilibrium states in a finite time, i.e. in an irreversible
way. Therefore, the irreversible counterparts of the classical heat
engines can be constructed at the mesoscopic level. In fact, our main
goal is building an irreversible version of the Carnot heat engine
with a colloidal particle in a harmonic trap of stiffness $k$,
immersed in a fluid at equilibrium with temperature $T$. This system
 is relevant from both the theoretical and experimental
standpoints. However, a difficulty arises: for mesoscopic systems, it
is impossible to completely decouple the system from the heat bath to
thermally isolate it---the interaction between a Brownian particle and
the fluid in which it is immersed cannot be switched off. Moreover,
zero heat and no entropy increment are not equivalent for finite time
processes.

The above discussion entails that the definition of adiabatic---in the
thermodynamical sense---process is far from trivial at the
mesoscale. Notwithstanding, very recently, \textit{finite-time
  adiabatic processes} have been characterised for a wide class of
mesoscopic systems~\cite{plata_finite-time_2020}, in the overdamped
description of the dynamics. In these processes, the average heat
vanishes for all times but there is entropy creation, as
imposed by the second principle. We employ these finite-time adiabatic
processes to build the corresponding adiabatic branches of the
irreversible Carnot engine.  Therefore, our approach differs from
other recent attempts to construct an irreversible Carnot
engine~\cite{schmiedl_efficiency_2008,bo_entropic_2013,tu_stochastic_2014,martinez_brownian_2016,nakamura_fast_2020},
the limitations of which are discussed in what follows. Specifically,
we focus on the respective definitions of ``adiabaticity''. In
Ref.~\cite{schmiedl_efficiency_2008}, working in the overdamped
regime, the term adiabatic has been employed for a process in which
the bath temperature $T$ is instantaneously changed, while the
configurational distribution is frozen.  However, as already noted by
the authors of that work, neither heat nor the entropy increment
vanishes in such a process, which are thus non-adiabatic, because of
the kinetic contribution thereto. In
Refs.~\cite{bo_entropic_2013,martinez_brownian_2016,nakamura_fast_2020},
the adiabatic branches are constructed by changing both the
temperature of the bath and the stiffness $k$ of the trap but keeping the
ratio $T^{2}/k$ constant, which is obtained in the underdamped
description. Nevertheless, the condition $T^{2}/k=\textnormal{const.}$
has been shown to correspond to isoentropic processes only in the
quasi-static limit~\cite{bo_entropic_2013,martinez_adiabatic_2015}, so
such a process is not adiabatic either for finite time
operation.~\footnote{Reference~\cite{plata_finite-time_2020}, although
  working in the overdamped description, incorporates the kinetic
  contribution to the energy balance. Therein, the ratio $T^{2}/k$ has
  been shown to be a non-decreasing function of time for finite-time
  adiabatic processes, being constant only in the quasi-static limit.}
A completely different approach is proposed in
Ref.~\cite{tu_stochastic_2014}. Therein, the oscillator follows a
Hamiltonian dynamics and is completely decoupled from the heat bath
during the adiabatic branches, a procedure that cannot be implemented
with a Brownian particle immersed in a fluid.


The performance of a heat engine is characterised by its efficiency
and power. The maximum efficiency achievable operating between a hot
bath at temperature $T_h$ and a cold bath at temperature $T_c$ is the
well-known Carnot efficiency $\eta_\Carnot=1-T_c/T_h$. However, it is
only reached for infinite time operation, which makes the power
vanish. The cycle must be swept in a finite time to yield a nonzero
power output. This acceleration of the process entails a
non-equilibrium dynamics and reduces the reachable efficiency. The
study of efficiency at maximum power is a classical problem associated
with the field of finite-time thermodynamics~\cite{curzon_efficiency_1975,andresen_thermodynamics_1977,
  de_vos_efficiency_1985,hoffmann_optimal_1985, chen_effect_1989,
  chen_maximum_1994}. Curzon and Ahlborn derived that the efficiency
at maximum power for a macroscopically endoreversible heat engine is
given by $\eta_\CA=1-\sqrt{T_c/T_h}$~\cite{curzon_efficiency_1975}.

There is no general proof ensuring that the efficiency of any
arbitrary heat engine at maximum power is bounded by the
Curzon-Ahlborn value. Nevertheless, myriads of different studies hint
at the existence of some universal properties, connected to the
Curzon-Ahlborn bound, for the efficiency at maximum
power. Specifically, it has been proven that in the limit of small
relative temperature difference, the two first terms in the expansion
of the efficiency at maximum power in the Carnot efficiency are
universal
\cite{van_den_broeck_thermodynamic_2005,esposito_universality_2009,
  sheng_constitutive_2015}.  This finding is completely consistent
with the results for the efficiency at maximum power in different
stochastic heat engines constructed either with a Brownian particle
\cite{schmiedl_efficiency_2008}, a Feynman ratchet
\cite{tu_efficiency_2008}, or a quantum dot
\cite{esposito_thermoelectric_2009}.

Another main objective of our work is the optimisation of the
irreversible Carnot engine. Specifically, in connection with the
discussion above, we are interested in looking into the optimisation
in a sense to be specified soon below,
of the delivered power and its associated efficiency. In this regard,
the optimal protocols for isothermal and the adiabatic branches, which
have been explicitly worked out
recently~\cite{schmiedl_optimal_2007,schmiedl_efficiency_2008,
  plata_optimal_2019,plata_finite-time_2020}, play a crucial
role. It appears that work should be minimised in the isothermal
processes~\cite{schmiedl_optimal_2007,schmiedl_efficiency_2008,
  plata_optimal_2019}, whereas the connection time is minimised in the
adiabatic ones~\cite{plata_finite-time_2020}.

The rest of the paper is organised as follows. In section
\ref{sec:definitions}, we introduce the model system with which we
construct our heat engine: a Brownian particle moving in a harmonic
trap. Special attention is paid to its energetics. Section
\ref{sec:isoth-adiabatic-processes} is devoted to putting forward the
optimal protocols for both the isothermal and adiabatic
branches. These protocols allow us to build the Carnot-like cycle,
which is analysed in section \ref{sec:irrev-Carnot-engine}. The
efficiency at maximum power is thoroughly investigated in section
\ref{sec:Curzon}. In section \ref{sec:conclusions}, the main
conclusions of our work are presented. Finally, we refer to the
Supplementary Material for some further technical details, which
complement the main text.

\section{The model system}
\label{sec:definitions}

\subsection{Definition}

We consider a one-dimensional (1D) overdamped harmonic oscillator of
stiffness $k$ in contact with a thermal bath at temperature $T$. 
A Brownian particle, confined by optical tweezers, provides an accurate realisation.
The
stochastic dynamics of the system may be modelled at either the
Langevin or the Fokker-Planck levels of description. The average
variance of the oscillator $\left\langle x^{2} \right\rangle$, which
is given by
\begin{equation}\label{eq:x2-overdamped}
  \lambda\frac{d\left\langle x^{2}\right\rangle}{dt}= - 2k \left\langle x^{2} \right\rangle +2 k_{B} T,
\end{equation}
being $k_{B}$ the Boltzmann constant and $\lambda$ the friction
coefficient. Physically, we are considering that the harmonic
oscillator is immersed in a certain fluid that plays the role of the
heat bath, which provides the values of the temperature $T$ and the
friction coefficient $\lambda$. Throughout
our work, we take $\lambda$ as constant\footnote{Indeed, 
when considering a colloidal particle in an optical trap, $\lambda$ is rooted in the solvent
viscosity and is essentially constant. }, time-independent, but we
assume both the stiffness of the oscillator $k$ and the bath
temperature $T$ to be externally controlled. 
While the time control of trap stiffness is now routinely achieved
experimentally, we refer to Refs. \cite{martinez_effective_2013,ciliberto_experiments_2017} for the time
control of temperature.

At any time $t$, the state of the system is characterised by the
state-point $(k,\langle x^{2}\rangle,T)$. Equilibrium states fulfil
the \textit{equation of state}
$\left\langle x^{2} \right\rangle_{\eq}=k_{B} T/k$.  The above
relation, which has been obtained by making the rhs of equation
\eref{eq:x2-overdamped} vanish, defines the equilibrium surface in the
$(k,\langle x^{2}\rangle,T)$ three-dimensional space. We want to
describe the energetics of this system at the average level. Thus we
define the average energy
\begin{equation}\label{eq:energy-def}
E=\frac{1}{2} k \left\langle x^2 \right\rangle +\frac{1}{2}k_BT,
\end{equation}
where we have taken into account that the velocity variable is always
at equilibrium in the overdamped limit.  The equilibrium value
of the energy is then $E_{\eq}=k_{B}T$.

Let us now consider a process starting from a certain state $A$ and
ending in another state $B$. Work and heat are defined by the
relations \cite{sekimoto_stochastic_2010}
\begin{equation}\label{eq:work-def}
W_{AB}=\frac{1}{2}\int_{A}^{B} \left\langle x^2 \right\rangle dk,
\end{equation}
\begin{equation}\label{eq:heat-def}
Q_{AB}=\frac{1}{2} \int_{A}^{B} \left( k \, d\!\left\langle x^2
  \right\rangle + k_B\, dT\right)= \frac{1}{2}\int_{A}^{B}  k \,
d\!\left\langle x^2 \right\rangle + \frac{k_B}{2}\,\left( T_{B}-T_{A}\right),
\end{equation}
where $T_{A}$ and $T_{B}$ are the temperature values for the initial
and final states, $A$ and $B$, respectively. Thus, the first law of
thermodynamics reads $\Delta E\equiv E_{B} - E_{A} = W_{AB}+Q_{AB}$.
We have used the following sign convention: for $W,Q>0$ energy is transferred from the environment to the system,
whereas for $W,Q<0$ energy is transferred from the system to the
environment, irrespective of the ``kind'' of energy
involved. Therefore, in order to consider a heat engine, we are
interested in cycles with a negative total work.

\subsection{Non-dimensional variables}

First of all, we introduce dimensionless variables as follows: we
divide the stiffness and the temperature by their respective initial
values, $\kappa=k/k_{0}$, $\theta=T/T_{0}$, and the variance by its
equilibrium value at the initial temperature,
$y=\left\langle x^2 \right\rangle/\left\langle x^2
\right\rangle_{\eq,0}=k_{0}\left\langle x^2 \right\rangle/(k_BT_{0})$.
Then, we have that $y(t=0)=1$ if the system starts from an equilibrium
state. Second, a dimensionless time is defined as $s=k_{0}t/\lambda$.
With the above definitions, the evolution of the system in
non-dimensional variables is governed by
\begin{equation}\label{eq:nd-y-overdamped}
\frac{dy}{ds}=-2\kappa y+2\theta,
\end{equation}
where the equilibrium surface (or equation of state) reads,
\begin{equation}\label{eq:y-eq}
\kappa y_{\eq}=\theta.
\end{equation}

Regarding the energetics, we introduce the dimensionless energy by
dividing $E$ by the equilibrium value at the initial time,
$k_B T_{0}$. Consistently, non-dimensional work and heat are defined
with the same energy unit, that is,
\begin{eqnarray} \label{eq:nd-energy-work-heat-def}
\mathcal{E}=\frac{1}{2} \kappa y +\frac{1}{2} \theta,
  \\
\label{eq:nd-work-heat-def}
\mathcal{W}_{AB}=\frac{1}{2}\int_{A}^{B} y \, d\kappa , \quad  
\mathcal{Q}_{AB}=
  \frac{1}{2}\int_{A}^{B} \kappa \, dy + \frac{1}{2}\left(\theta_{B}-\theta_{A}\right).
\end{eqnarray}
The first law reads $\Delta\mathcal{E}\equiv\mathcal{E}_{B} - \mathcal{E}_{A} = \mathcal{W}_{AB}+\mathcal{Q}_{AB}$,
and the equilibrium value of the energy is $\mathcal{E}_{\eq}=\theta$.

In dimensionless variables, the state of the system is characterised
by the state-point $(\kappa,y,\theta)$ at any time $s$. For our
purposes, it is useful to consider the movement of the projection of
the state-point onto the $(\kappa,y)$ plane. In particular, the work
$\mathcal{W}$, as given by equation \eref{eq:nd-work-heat-def}, is proportional
to the area below the curve $(\kappa(s),y(s))$ swept by the system as
time increases.

\section{Building blocks of the cycle: Isothermal and adiabatic processes}
\label{sec:isoth-adiabatic-processes}

Herein, we aim at building an irreversible heat engine with the above
described overdamped harmonic oscillator.  Our heat engine operates
cyclically between a ``hot'' source, at dimensionless temperature
$\theta_{h}$, and a ``cold'' source, at temperature
$\theta_{c}<\theta_{h}$. Specifically, the non-equilibrium cycle
comprises four different processes: two isothermal ones, at
temperatures $\theta_{h}$ and $\theta_{c}$, and two \textit{locally}
adiabatic ones that connect the isotherms. No heat is exchanged in
average during these locally adiabatic processes at all times, as described
below. This is the usual use of the term adiabatic in equilibrium
thermodynamics, in which adiabatic is employed for a process in which
the system is thermally insulated from the
environment.

In each cycle, the engine takes energy from the hot
reservoir as heat, $\mathcal{Q}_{h}>0$, and performs work, that is,
$\mathcal{W}<0$. Therefore, the projection of the state-point onto the
$(\kappa,y)$ plane sweeps a certain closed curve $(\kappa(s),y(s))$,
which characterises the considered cycle, in the counterclockwise
direction.
In the light of the above, isothermal and adiabatic processes
can be considered as the \textit{building blocks} for our irreversible
heat engine. In the following, we summarise some results
obtained in previous studies for isothermal \cite{schmiedl_efficiency_2008,plata_optimal_2019} and adiabatic processes \cite{plata_finite-time_2020}.

\subsection{Isothermal processes}\label{sec:isoth-process}

We consider two kinds of isothermal processes at temperature $\theta$:
quasi-static and optimal. In both of them, the initial and final states
characterised by $(\kappa_{A},y_{B})$ and
$(\kappa_{A},y_{B})$, respectively, correspond to equilibrium
situations. Therefore, $\kappa_{A}y_{A}=\kappa_{B}y_{B}=\theta$.

First, we deal with the quasi-static case. Therein, $\kappa$ is slowly
tuned in such a way that the system sweeps the equilibrium curve $y(s)=\theta/\kappa(s)$
in the $(\kappa,y)$ plane. Therefrom,
\begin{equation}
\mathcal{W}=\frac{\theta}{2} \ln \frac{\kappa_{B}}{\kappa_{A}},
\quad
\mathcal{Q}=-\mathcal{W},
\quad 
\Delta E=0, \quad \mathcal{E}_{B}=\mathcal{E}_{A}=\theta.
\end{equation}
Of course, this quasi-static process takes an infinite time.

Second, we look into the optimal process for a given finite time
$s_{\fin}$. Therein, we are interested in the process for which the
work performed by an external agent on the system is minimum, or in other words, we look for the maximum work produced by the system. The
evolution of the variance in the 
optimal process is \cite{schmiedl_efficiency_2008,plata_optimal_2019}.
\begin{equation}\label{eq:y-optimal}
\tilde{y}(s)= \left[ \sqrt{y_{A}} + \left( \sqrt{y_{B}} -\sqrt{y_{A}} \right) \frac{s}{s_{\fin}} \right]^2.
\end{equation}
From now, tilde denotes optimality in some sense: either for the
profiles or for the values of the physical quantities or parameters. Note
that $\tilde{y}(s)$ is continuous in the whole interval
$[0,s_{\fin}]$.

The optimal evolution for the stiffness is obtained from the evolution
equation \eref{eq:nd-y-overdamped} in the open interval
$(0,s_{\fin})$,
\begin{equation}\label{eq:kappa-optimal}
\tilde{\kappa}(s)= \frac{\theta}{\tilde{y}(s)}-\frac{1}{2}\frac{d}{ds}\ln \tilde{y}(s), \quad
0<s<s_{f}.
\end{equation}
We recall that the stiffness is discontinuous at both the initial and
final times, $\tilde{\kappa}(s=0)=\kappa_{A}$, $\tilde{\kappa}(s=s_{\fin})=\kappa_{B}$.
In this problem, the elastic constant $\kappa(s)$ plays the role of
the ``control'' function in optimal control
theory~\cite{pontryagin_mathematical_1987,liberzon_calculus_2012}. Similar
discontinuities in the ``control'' function have been repeatedly found
in stochastic thermodynamics
\cite{schmiedl_efficiency_2008,plata_optimal_2019,band_finite_1982,schmiedl_optimal_2007,aurell_optimal_2011,aurell_boundary_2012,muratore-ginanneschi_application_2017}.\footnote{This
  is a consequence of the corresponding ``Lagrangian'' being linear in
  the ``velocities'' \cite{newman_lagrangians_1955}, which is
  sometimes called the Miele problem \cite{tolle_optimization_2012}.}

The optimal values of work and heat can also be readily
calculated. The results are 
\begin{eqnarray}
\widetilde{\mathcal{W}}=\frac{\theta}{2} \ln \frac{\kappa_{B}}{\kappa_{A}} + \frac{\theta}{s_{\fin}} \left( \frac{1}{\sqrt{\kappa_{B}}}-\frac{1}{\sqrt{\kappa_{A}}} \right)^2,
\quad
\widetilde{\mathcal{Q}}= -\widetilde{\mathcal{W}},
\end{eqnarray}
Of course, in this isothermal process there is no energy change
between the initial and final states $\Delta\mathcal{E}=0, \quad \mathcal{E}_{B}=\mathcal{E}_{A}=\theta$.
Note, however, that the energy of the system does change in the
intermediate times, $\mathcal{E}(s)\neq\theta$ for $0<s<s_{\fin}$
because we are dealing with a non-equilibrium process and
$y(s)\neq \theta/\kappa(s)$, as expressed by equation \eref{eq:kappa-optimal}.

\subsection{Adiabatic processes}\label{sec:adiab-process}

Now we turn our attention to adiabatic processes, there is no heat
transfer at any point of the system trajectory. Therefore, bearing in
mind equation \eref{eq:nd-work-heat-def} we have that the
infinitesimal heat vanishes, i.e. 
\begin{equation}\label{eq:adiabatic-def}
\dbar Q\equiv \kappa \, dy + d\theta = 0,
\end{equation}
Note that temperature becomes a function of time that goes from
$\theta_{A}$ to $\theta_{B}$ in adiabatic processes. Similarly
to the case of isothermal processes, we only consider adiabatic
processes connecting two equilibrium states and then $\kappa_{A}
y_{A}=\theta_{A}$, $\kappa_{B} y_{B}=\theta_{B}$.

The energetics of adiabatic processes is quite simple. The energy
change is given by the change in temperature,
$\mathcal{E}_{A}=\theta_{A}$, $\mathcal{E}_{B}=\theta_{B}$,
$\Delta\mathcal{E}=\theta_{B}-\theta_{A}$.
Since there is no heat exchange, $\mathcal{Q}= 0$, 
work coincides with the energy change,
$\mathcal{W}=\theta_{B}-\theta_{A}$. 
The above expressions for energy, heat and work apply for any
adiabatic process, regardless of its duration, and therefore are valid
for both quasi-static and non-equilibrium processes. Nevertheless, the equivalence between adiabatic and isoentropic processes occurs only in the quasi-static limit. It is in the non-equilibrium case that we deviate from the proposals in Refs.~\cite{martinez_brownian_2016, bo_entropic_2013}.

Again we consider two kinds of  processes: quasi-static and
optimal. First, in the quasi-static case, $\kappa$ and $\theta$ are
tuned in an infinitely slow way to allow the system sweep the
equilibrium curve \eref{eq:y-eq}. Combining equations \eref{eq:y-eq} and
\eref{eq:adiabatic-def}, one gets
\begin{equation}\label{eq:adiabatic-quasistatic}
y(s)=\frac{y_{A}\theta_{A}}{\theta(s)}=y_{A} \sqrt{\frac{\kappa_{A}}{\kappa(s)}}.
\end{equation}
Second, we investigate optimal adiabatic processes. Here, optimal
means something different from the sense we used in the previous
section. As already said above, the work value is fixed by the initial
and target states and thus cannot be optimised. However, two arbitrary
states cannot be connected by an adiabatic transformation, the following inequality
\begin{equation}\label{eq:adiabatic-inequality-all-times}
\frac{\theta(s)}{\theta_{A}}\geq\left(\frac{y(s)}{y_{A}}\right)^{-1},
\end{equation}
holds for all times~\cite{plata_finite-time_2020}. Therefore, for the initial and final times,
\begin{equation}\label{eq:adiabatic-inequality} 
  \frac{\theta_{B}}{\theta_{A}}\geq
  \left(\frac{y_{B}}{y_{A}}\right)^{-1}
\textnormal{ or, equivalenty, } \left(\frac{\theta_{B}}{\theta_{A}}\right)^2\geq\frac{\kappa_{B}}{\kappa_{A}}
\end{equation}
must be fulfilled. The equality in equations
\eref{eq:adiabatic-inequality-all-times} and
\eref{eq:adiabatic-inequality} corresponds to the quasi-static case
\eref{eq:adiabatic-quasistatic}.

There exists a minimum time to carry out an adiabatic
process \cite{plata_finite-time_2020}, namely
\begin{equation}\label{eq:smin-adiab}
\tilde{s}_{\fin}=\frac{\left(y_{B}-y_{A}\right)^2}{2\left(y_{B}\theta_{B}-y_{A}\theta_{A}\right)}.
\end{equation}
This minimum time is reached for a protocol in which the variance and
the temperature 
evolve according to
\begin{equation}\label{eq:y-theta-evol-adiab}
\tilde{y}(s)=y_{A}+(y_{B}-y_{A})\frac{s}{\tilde{s}_{\fin}}, \quad \tilde{\theta}(s)=\frac{y_{A}\theta_{A}+(y_{B}\theta_{B}-y_{A}\theta_{A})\frac{s}{\tilde{s}_{\fin}}}{y_{A}+(y_{B}-y_{A})\frac{s}{\tilde{s}_{\fin}}},
\end{equation}
which are valid in the whole interval
$[0,\tilde{s}_{\fin}]$. Therefore, both $\tilde{y}(s)$ and
$\tilde{\theta}(s)$ are continuous functions of time, including the
initial and final times. The stiffness is given by
\begin{equation}\label{eq:kappa-evol-adiab}
\tilde{\kappa}(s)= -\left({\frac{d\tilde{y}(s)}{ds}}\right)^{-1}\frac{d\tilde{\theta}(s)}{ds} , \quad
0<s<\tilde{s}_{f}.
\end{equation}
and $\tilde{\kappa}(s=0)=\kappa_{A}$,  $\tilde{\kappa}(s=\tilde{s}_{\fin})=\kappa_{B}$.

The discontinuity at the boundaries of $\kappa(s)$ does not break the
adiabatic character of the process: there is no
\textit{instantaneous} heat transfer at the initial and/or final
times. Since both the variance $y$ and the temperature $\theta$ are
continuous at the boundaries, the integration of the differential of
heat, as defined in equation~\eref{eq:adiabatic-def}, between $s=0$ and
$s=0^{+}$ (or between $s=\tilde{s}_{f}^{-}$ and $s=\tilde{s}_{f}$) vanishes. On the
contrary, there is an instantaneous contribution to the work at both
boundaries.

\section{Irreversible Carnot-like heat engine}
\label{sec:irrev-Carnot-engine}

The aim of this work is to study a (stochastic) thermodynamic cycle
comprising the following processes: (i) Isothermal expansion starting
from $(\kappa_A,y_A,\theta_A)$ up to
$(\kappa_B,y_B,\theta_B=\theta_{A})$ in contact with a hot bath at
temperature $\theta_{A}$, (ii) adiabatic expansion starting from
$(\kappa_B,y_B,\theta_B=\theta_{A})$ up to $(\kappa_C,y_C,\theta_C)$,
(iii) isothermal compression starting from $(\kappa_C,y_C,\theta_C)$
up to $(\kappa_D,y_D,\theta_D=\theta_{C})$ in contact with a cold bath
at temperature $\theta_{C}$, and (iv) adiabatic compression going from
$(\kappa_D,y_D,\theta_D=\theta_{C})$ to $(\kappa_A,y_A,\theta_A)$.
We always choose the normalisation constants (units) such that
$(\kappa_A,y_A,\theta_A)=(1,1,1)$.

As a consequence of the above processes being isothermal/adiabatic, we
have the following general identities,
$\mathcal{W}_{AB}=-\mathcal{Q}_{AB}$,
$\mathcal{W}_{BC}=\mathcal{E}_{C}-\mathcal{E}_{B}=\theta_{C}-\theta_{A}$,
$\mathcal{Q}_{BC}=0$, $\mathcal{W}_{CD}=-\mathcal{Q}_{CD}$,
$\mathcal{W}_{DA}=\mathcal{E}_{A}-\mathcal{E}_{D}=\theta_{A}-\theta_{C}=-\mathcal{W}_{BC}$,
$\mathcal{Q}_{DA}=0$.
We focus on a heat engine, that is, a device that extracts heat from
the hot bath and performs work, i.e.
\begin{equation}\label{eq:heat-engine-condition}
  \mathcal{Q}_{AB}=-\mathcal{W}_{AB}>0, \quad \mathcal{W}_{AB}+\cancel{\mathcal{W}_{BC}}+\mathcal{W}_{CD}+\cancel{\mathcal{W}_{DA}}=\mathcal{W}_{AB}+\mathcal{W}_{CD}<0.
\end{equation}
The efficiency of such a device is defined by
\begin{equation}\label{eq:efficiency-def}
\eta\equiv\frac{-\left(\mathcal{W}_{AB}+\mathcal{W}_{CD}\right)}{\mathcal{Q_{AB}}}=1-\frac{\mathcal{W}_{CD}}{\mathcal{Q}_{AB}}<1,
\end{equation}
whereas the power that delivers is given by
\begin{equation}\label{eq:power-def}
  \mathcal{P}\equiv\frac{-\left(\mathcal{W}_{AB}+\mathcal{W}_{CD}\right)}{s_{AB}+s_{BC}+s_{CD}+s_{DA}},
\end{equation}
where $s_{AB}$ is the time employed for going from $A$ to $B$, and so on.

\subsection{Quasi-static case}\label{sec:quasistatic}

First, we concentrate on the quasi-static limit, that is, we consider a
Carnot cycle in which the harmonic oscillator is always at
equilibrium.  In principle, we must give 12 numbers to characterise
the four operating points of the cycle $(A,B,C,D)$, but we have the
following constraints: (i) due to normalisation, state $A$ is given,
$(\kappa_A,y_A,\theta_A)=(1,1,1)$ (3 constraints), (ii) points
$(B,C,D)$ are equilibrium states (3 constraints), (iii) two isothermal
relations $A$-$B$ and $C$-$D$ (2 constraints), and (iv) two adiabatic
relations $B$-$C$ and $D$-$A$ (2 constraints).
So, we need only $12-3-3-2-2=2$ variables to univocally define the
quasi-static cycle.

The cycle is thus completely characterised by the temperature ratio
$\nu$ and the compression ratio along the first isotherm $\chi$. The
values of the state variables $(\kappa,y,\theta)$ at the operating
points of the cycle are collected in panel (a) of
Table~\ref{tab:Carnot}. Note that the isotherm condition implies that
$y_{B}/y_{A}=\kappa_{A}/\kappa_{B}=\chi^{-1}$, so the parameter $\chi$
certainly gives the compression ratio along the first
isotherm. Hereafter, to keep our wording simpler, we call $\chi$ the
compression ratio.
\begin{table}
\centering
\caption{\label{tab:Carnot} Operating points of the
  Carnot engines. Panels (a) and (b) correspond to the reversible and
  irreversible versions, respectively.}
\vspace{.1cm}
\begin{tabular}{p{1cm}|p{1cm}|p{1.25cm}|p{0.5cm}|}
\cline{2-4}
          (a)                & $\kappa$ & $y$ & $\theta$ \\ \hline
\multicolumn{1}{|c|}{$A$} & $1$ & $1$ & $1$ \\ \hline
\multicolumn{1}{|c|}{$B$}  & $\chi$ & $\chi^{-1}$ & $1$ \\ \hline
\multicolumn{1}{|c|}{$C$}    & $\nu^2 \chi$ & $\nu^{-1}\chi^{-1}$ & $\nu$ \\ \hline
\multicolumn{1}{|c|}{$D$}    & $\nu^2$ & $\nu^{-1}$ & $\nu$ \\ \hline
\end{tabular} \hspace{1cm}
\begin{tabular}{p{1cm}|p{1cm}|p{1.75cm}|p{0.5cm}|}
\cline{2-4}
       (b)                   & $\kappa$ & $y$ & $\theta$ \\ \hline
\multicolumn{1}{|c|}{$A$} & $1$ & $1$ & $1$ \\ \hline
\multicolumn{1}{|c|}{$B$}  & $\chi$ & $\chi^{-1}$ & $1$ \\ \hline
\multicolumn{1}{|c|}{$C$}    & $c \nu^2 \chi$ & $c^{-1}\nu^{-1}\chi^{-1}$ & $\nu$ \\ \hline
\multicolumn{1}{|c|}{$D$}    & $d \nu^2$ & $d^{-1}\nu^{-1}$ & $\nu$ \\ \hline
\end{tabular}
\end{table}

The efficiency of a Carnot cycle
\begin{equation}
\label{Carnot_eff}
\eta_{\Carnot}=1-\frac{\theta_C}{\theta_A}=1-\nu,
\end{equation}
is well known and can be derived for any system without the knowledge
of its state equation through entropic considerations
\cite{callen_thermodynamics_1985}. Here, it can also be explicitly checked
by calculating work and heat over the branches of the cycle. 
The power delivered by this engine is zero, because the processes
are quasi-static and thus involve an infinite time.

\subsection{Irreversible Carnot-like cycle at finite speed}\label{sec:Carnot-finite-speed}

\begin{figure}
   \includegraphics[width= \textwidth]{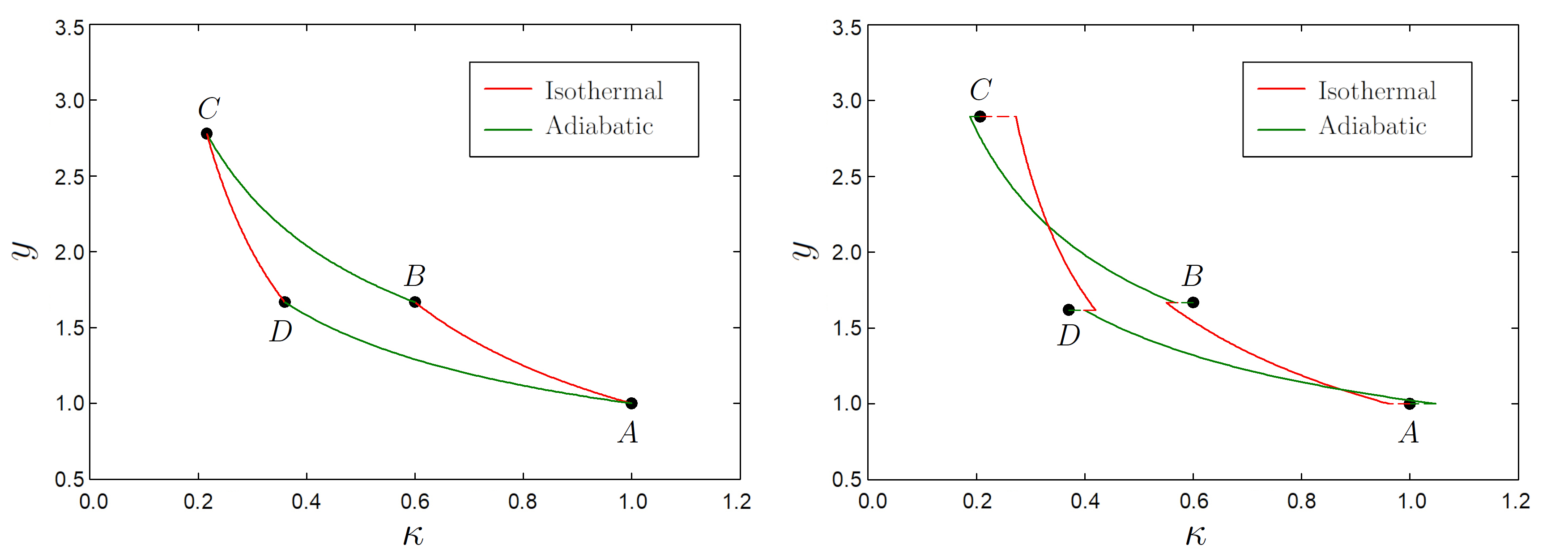}
  \caption{\label{fig:Carnot-engines} (a) Projection of the movement
    of the state-point onto the $(\kappa,y)$ plane for a reversible
    Carnot engine. (b) Projection of the movement of the state-point
    onto the $(\kappa,y)$ plane for an irreversible Carnot-like
    engine. Specifically, we have used the parameter values $\nu=0.6$,
    $\chi=0.6$, $c=0.96$, $d=1.03$ and the corresponding optimal protocols
    discussed in the text. In both plots, red lines correspond to the
    isothermal processes and green lines to the adiabatic ones. The
    dashed segments mark the jumps in the stiffness at the initial and
    final points of each of the four branches of the cycle.}
\end{figure}

Now we consider a similar cycle, being the only difference that the
processes are carried out in a finite time and are thus
irreversible. The adiabaticity of the second and third process impose
two inequalities, as expressed by equation
\eref{eq:adiabatic-inequality}. Therefore, we have that
\begin{equation}\label{eq:ineq-adiab}
  \frac{\theta_C^2}{\theta_B^2} \geq \frac{\kappa_C}{\kappa_B}, \qquad
  \frac{\theta_A^2}{\theta_D^2} \geq \frac{\kappa_A}{\kappa_D},
\end{equation}
which become equalities only for reversible processes, as those in the
previous section. Thus, we need two additional parameters to define
the cycle unambiguously, specifically we choose to introduce
\begin{equation}\label{eq:c-and-d-def}
  c=\kappa_C \nu^{-2} \chi^{-1}\leq 1, \qquad d=\kappa_D \nu^{-2}\geq 1,
\end{equation}
which assure that equation \eref{eq:ineq-adiab} is fulfilled. In panel
(b) of Table~\ref{tab:Carnot}, we summarise the values of the state
variables $(\kappa,y,\theta)$ at the operating points of this
non-equilibrium cycle. A comparative plot of the reversible and
irreversible Carnot engines is shown in
figure~\ref{fig:Carnot-engines}.


In the following, we focus our attention on the maximisation of the
power delivered by the engine. Therefore, we build
the heat engine that operates at maximum power for fixed operating
points $(A,B,C,D)$ or, equivalently, for given values of
$(\nu,\chi,c,d)$.  We approach the problem of the maximisation of the power defined
in equation \eref{eq:power-def} in a stepwise manner. 
As discussed in detail below,
the main idea is that
the global maximum of $\mathcal{P}$ can be obtained as the maximum of
maximums, that is, we start by maximising with respect to some
parameters keeping the remainder fixed. Afterwards, this maximum can
be in turn be maximised with respect to the previously fixed
parameters. For instance, maximisation can be performed for 
a given set $(\nu,\chi,c$ and $d$), or in a more global fashion, specifying
$\nu$ only.

Maximising equation \eref{eq:power-def} implies to take the shortest
possible adiabatic protocols and the minimal work for the isothermal
processes.  This is readily understood as follows. The only dependence
on the adiabatic protocols come from $s_{BC}$ and $s_{DA}$, so they
have to be minimum in order to give the maximum value for
$\mathcal{P}$. With respect to the isothermal processes, for fixed
values of $s_{AB}$ and $s_{CD}$, we have to maximise the respective
work values $-\mathcal{W}_{AB}$ and $-\mathcal{W}_{CD}$, that is,
minimise $\mathcal{W}_{AB}$ and $\mathcal{W}_{CD}$.  Therefore, we end
up with the \textit{optimal} processes, either isothermal or
adiabatic, discussed in section
\ref{sec:isoth-adiabatic-processes}. Making use of equations
\eref{eq:y-optimal}-\eref{eq:kappa-optimal} for the isothermal
processes and equations
\eref{eq:smin-adiab}-\eref{eq:kappa-evol-adiab} for the adiabatic
ones, we get
\begin{eqnarray}
  \mathcal{W}_{AB}=\frac{1}{2}\ln \chi + \frac{1}{s_{AB}}\left( \frac{1}{\sqrt{\chi}}-1\right)^2,\quad \mathcal{Q}_{AB}=-\mathcal{W}_{AB},
  \\ 
  \mathcal{W}_{CD}=-\frac{\nu}{2} \ln \frac{c \chi}{d}+ \frac{1}{\nu s_{CD}}\left( \frac{1}{\sqrt{c \chi}}-\frac{1}{\sqrt{d}}\right)^2 ,\quad \mathcal{Q}_{CD}=-\mathcal{W}_{CD} ,
  \\ 
\label{eq:sBC_sDA_opt}
\tilde{s}_{BC}=\frac{(1-c\nu)^2}{2c\chi \nu^2(1-c)}, \quad \tilde{s}_{DA}=\frac{(d\nu-1)^2}{2d \nu^2(d-1)}
\end{eqnarray}

It is worth commenting some points before proceeding further. On the
one hand, $\mathcal{W}_{CD}$ is always positive and thus
$\mathcal{Q}_{CD}$ is negative; isothermal compression work has
to be done on the system and heat is always transferred from the
device to the cold bath. On the other hand, in the isothermal
expansion, $\mathcal{W}_{AB}$ is negative for large enough $s_{AB}$,
but $\mathcal{W}_{AB}$ becomes positive if we intend to compress the
system too fast: we have to exert work on the system in that case and
moreover $Q_{AB}$ becomes negative and heat is transferred from the
system to the hot bath. Therefore, we are not interested here in these
too fast isothermal expansions because we would not be building a heat
engine in that case. Below we show that this poses no problem because
(i) the optimal value $\tilde{s}_{AB}$ yields a negative value of
$\mathcal{W}_{AB}$ and (ii) the optimal value $\tilde{s}_{CD}$ makes
that $\mathcal{W}_{CD}<-\mathcal{W}_{AB}$, that is,
$\mathcal{W}_{CD}+\mathcal{W}_{AB}<0$. Thus, the heat engine
conditions are met.

Let us build on the ideas above.  We must impose the inequalities
\eref{eq:heat-engine-condition} to have a heat engine. In particular,
these inequalities should hold when $s_{AB}$ and $s_{CD}$ go to
infinity (reversible isotherms). It is useful to introduce the
definitions
\begin{equation}\label{eq:some-definitions}
\mathcal{W}_{1}\equiv\lim_{s_{AB}\to\infty}\mathcal{W}_{AB} =\frac{1}{2}\ln
  \chi <0, \quad
  \mathcal{W}_{2}\equiv\lim_{s_{CD}\to\infty}\mathcal{W}_{CD}
  =-\frac{\nu}{2}\ln \frac{c \chi}{d} >0,
\end{equation}
where we have taken into account that $\chi<1$, $c\leq 1$, $d\geq 1$,
and 
\begin{equation}
\label{eq:W-infty}
  \mathcal{W}_{\infty}\equiv\mathcal{W}_{1}+\mathcal{W}_{2}=\frac{1}{2}\ln
  \chi-\frac{\nu}{2}\ln \frac{c \chi}{d}<0.
\end{equation}
Although $\mathcal{W}_{1}$ coincides with the value of the work over
the first isotherm in the fully reversible engine, neither
$\mathcal{W}_{2}$ nor $\mathcal{W}_{\infty}$ does because they depend
on $c$ and $d$.  The negativeness of $\mathcal{W}_{\infty}$ leads to
the constraint
\begin{equation}\label{eq:c-d-chi-constraint}
\frac{c}{d}>\chi^{\frac{1-\nu}{\nu}}.
\end{equation}
Strictly speaking, this constraint has been shown to hold only in the limit as
$s_{AB},s_{CD}\to \infty$, but below we prove that it also holds for
finite-time operation. 

Therefore, we just have to maximise the power
\begin{equation}\label{eq:P-max-step2}
\mathcal{P}=\frac{\frac{\nu-1}{2} \ln \chi + \frac{\nu}{2} \ln \frac{c}{d} - \frac{1}{s_{AB}}\left(\frac{1}{\sqrt{\chi}}-1\right)^2-\frac{1}{\nu s_{CD}} \left(\frac{1}{\sqrt{d}}-\frac{1}{\sqrt{c \chi}}\right)^2}{s_{AB}+\tilde{s}_{BC}+s_{CD}+\tilde{s}_{DA}},
\end{equation}
with respect to $s_{AB}$ and $s_{CD}$, by imposing that the partial
derivatives of $\mathcal{P}$ with respect to $s_{AB}$ and $s_{CD}$
vanish for the optimal durations of the isothermal processes.\footnote{Note that, since they do not
depend on $s_{AB}$ and $s_{CD}$, we have not substituted explicitly
the values of $\tilde{s}_{BC}$ and $\tilde{s}_{DA}$, given by
equation \eref{eq:sBC_sDA_opt}, so as not to clutter the
expression.}  In order to write the expressions for $\tilde{s}_{AB}$
and $\tilde{s}_{CD}$, it is convenient to introduce the parameters
\begin{eqnarray}
  \Delta_1=\sqrt{y_B}-\sqrt{y_A}=\frac{1}{\sqrt{\chi}}-1>0,\\
\Delta_2=\sqrt{y_D}-\sqrt{y_C}=\frac{1}{\sqrt{\nu}}\left(
  \frac{1}{\sqrt{d}}-\frac{1}{\sqrt{c \chi}}\right)<0,
\end{eqnarray}
which  measure the expansion and
compression of the system in the first and second isotherms,
respectively, and
\begin{eqnarray}
\label{eq:sigma-definition}
\sigma=\sqrt{1+\frac{\left(\tilde{s}_{BC}+\tilde{s}_{DA}\right)\left(-\mathcal{W}_{\infty}\right)}{\left(\Delta_{1}-\Delta_{2}\right)^{2}}}>1.
\end{eqnarray}
 As a function of these parameters, we can write now that
\begin{eqnarray}
\label{eq:s-opt-iso1}
\tilde{s}_{AB}=
                   \frac{\Delta_{1}(\Delta_{1}-\Delta_{2})(1+\sigma)}
                   {-\mathcal{W}_{\infty}}, \quad
  \tilde{s}_{CD}=\frac{-\Delta_{2}(\Delta_{1}-\Delta_{2})(1+\sigma)}
  {-\mathcal{W}_{\infty}}.
\end{eqnarray}
The condition $\mathcal{W}_{\infty}<0$ 
ensures the positivity of the optimal times.

Using the above definitions,  we can write the work values for the
optimal durations of the isothermal processes as
\begin{eqnarray}
\label{eq:optimal-works-isotherms}
\widetilde{\mathcal{W}}_{AB}=\mathcal{W}_{1}+\frac{\Delta_{1}^{2}}{\tilde{s}_{AB}}
      =\frac{-\mathcal{W}_{1}\Delta_{2}-\mathcal{W}_{2}
                        \Delta_{1}+\mathcal{W}_{1}\sigma(\Delta_{1}-\Delta_{2})}{(\Delta_{1}-\Delta_{2})(1+\sigma)}<0, \\
\widetilde{\mathcal{W}}_{CD}=\mathcal{W}_{2}+\frac{\Delta_{2}^{2}}{\tilde{s}_{CD}}
      = \frac{\mathcal{W}_{1}\Delta_{2}+\mathcal{W}_{2}
       \Delta_{1}+\mathcal{W}_{2}\sigma(\Delta_{1}-\Delta_{2})}
       {(\Delta_{1}-\Delta_{2})(1+\sigma)}>0.
\end{eqnarray}
By combining the expressions above, the total work in the cycle with
the optimal durations is found to be
\begin{equation}
\widetilde{\mathcal{W}}_{AB}+\widetilde{\mathcal{W}}_{CD}=\mathcal{W}_{\infty}-\frac{1}{1+\sigma}
  \mathcal{W}_{\infty}=\frac{\sigma}{1+\sigma}W_{\infty}<0
\end{equation}
The signs of $\widetilde{\mathcal{W}}_{AB}$ and
$\widetilde{\mathcal{W}}_{AB}+\widetilde{\mathcal{W}}_{CD}$ show that
we have, in fact, a ``good'' engine. Moreover, we get a physical
interpretation for the parameter $\sigma$: it measures the deviation
of the total irreversible work from the value for infinitely slow
isothermal processes $\mathcal{W}_{\infty}$. In the limit as
$\sigma\to\infty$, we have that
$\widetilde{\mathcal{W}}_{AB}+\widetilde{\mathcal{W}}_{CD}\to
\mathcal{W}_{\infty}$.

We have found the optimal values of the times for the isothermal and
adiabatic protocols, for given values of the parameters
$(\nu,\chi,c,d)$ that univocally define the operating points of our
irreversible Carnot-like heat engine. As a function of these
parameters, the optimal power is thus given
by
\begin{equation}\label{eq:optimal-power}
  \widetilde{\mathcal{P}}=\frac{-\mathcal{W}_{\infty}-
    \frac{\Delta_{1}^{2}}{\tilde{s}_{AB}}-
    \frac{\Delta_{2}^{2}}{\tilde{s}_{CD}}}{\tilde{s}_{AB}+\tilde{s}_{BC}+\tilde{s}_{CD}+\tilde{s}_{DA}}
=\frac{-\mathcal{W}_{\infty}\,\frac{\sigma}{1+\sigma}}{\tilde{s}_{AB}+\tilde{s}_{BC}+\tilde{s}_{CD}+\tilde{s}_{DA}},
\end{equation}
Later, we address the issue of optimising the cycle further, by
looking for the maximum of the $\widetilde{\mathcal{P}}$ as a function
of $c$, $d$ and $\chi$ for a fixed value of the temperature ratio
$\nu$.

\section{Efficiency at maximum power and the Curzon-Ahlborn
  bound}\label{sec:Curzon}

\subsection{Maximal power at fixed temperature and compression ratios}

Let us look into the the efficiency of the maximum power cycle,
\begin{equation}
\label{eq:efficiency-at-optimal-power}
\tilde{\eta}=-\frac{\widetilde{\mathcal{W}}_{AB}+
  \widetilde{\mathcal{W}}_{CD}}{\widetilde{\mathcal{Q}}_{AB}}=1+
\frac{\widetilde{\mathcal{W}}_{CD}}{\widetilde{\mathcal{W}}_{AB}},
\end{equation}
which depends on $(\nu,\chi,c,d)$. To keep our notation
simple, either for $\widetilde{\mathcal{P}}$ in equation
\eref{eq:optimal-power} or $\tilde{\eta}$ in equation
\eref{eq:efficiency-at-optimal-power}, we do not write explicitly the
parameters which they depend on. This choice also applies to the
remainder of the paper.

Making use of equation \eref{eq:optimal-works-isotherms}, we can rewrite
$\tilde{\eta}$ as
\begin{eqnarray}\label{eq:tilde-eta-1}
  \tilde{\eta}&=
                & \underbrace{1-\nu}_{\eta_{\Carnot}}+\frac{(\nu-1)\left(\mathcal{W}_1\Delta_2 +
                  \mathcal{W}_2 \Delta_1\right)-\left(\mathcal{W}_2+\mathcal{W}_{1}\nu\right) \sigma
                  \left(\Delta_1-\Delta_2\right)}{\mathcal{W}_1 \Delta_2 +
                  \mathcal{W}_2 \Delta_1-\mathcal{W}_1 \sigma
                  \left(\Delta_1-\Delta_2\right)}
\end{eqnarray}
All the terms in the denominator are clearly positive, whereas all the
terms in the numerator are negative by taking into account that $\mathcal{W}_2+\mathcal{W}_{1}\nu=-(\nu/2) \ln(c/d)>0$.
Therefore, $\tilde{\eta}<\eta_{\Carnot}$: the efficiency is always below the Carnot bound, as expected.

On the other hand, the comparison with the Curzon-Ahlborn bound
\cite{curzon_efficiency_1975,schmiedl_efficiency_2008,esposito_efficiency_2010,apertet_true_2017}
\begin{equation}\label{eq:efficiency-CA}
  \eta_{\CA}=1-\sqrt{\nu}
\end{equation}
requires a more detailed analysis. Let us investigate two different
cases. First, we consider values of the ratio $c/d$ such that
$\chi^{-1+\nu^{-1}}<c/d<\chi^{-1+\nu^{-1/2}}$,
which entails that $\mathcal{W}_2+\sqrt{\nu} \mathcal{W}_1>0$.
for which
\begin{equation}
\underbrace{\left(\sqrt{\nu}-1\right)}_{<0}(\underbrace{\mathcal{W}_1 \Delta_2}_{>0} + \underbrace{\mathcal{W}_2 \Delta_1}_{>0})-\underbrace{(\mathcal{W}_2+ \mathcal{W}_1\sqrt{\nu})}_{>0}\sigma \underbrace{(\Delta_1 -\Delta_2)}_{>0}<0.
\end{equation}
In this region, a manipulation similar to the one done for showing
that $\eta<\eta_{\Carnot}$ gives
\begin{equation}
 \tilde{\eta}=1-\sqrt{\nu}+\frac{(\sqrt{\nu}-1)\left(\mathcal{W}_1\Delta_2 +
        \mathcal{W}_2 \Delta_1\right)-\left(\mathcal{W}_2+\mathcal{W}_{1}\sqrt{\nu}\right) \sigma
        \left(\Delta_1-\Delta_2\right)}{\mathcal{W}_1 \Delta_2 +
        \mathcal{W}_2 \Delta_1-\mathcal{W}_1 \sigma
        \left(\Delta_1-\Delta_2\right)}.
\end{equation}
Again, the denominator and the numerator are positive and negative
respectively, which leads to the inequality $\tilde{\eta}<\eta_{CA}$.
Nevertheless, for the complementary case, $\chi^{-1+\nu^{-1/2}}<c/d<1$,
we can no longer assure that the Curzon-Ahlborn is an upper
bound. Indeed, in the double limit as $(c,d)\to(1,1)$, we have that
$\tilde{s}_{BC}$ and $\tilde{s}_{DA}$ diverge for fixed $\nu<1$. In
that limit, not only do the adiabatic processes become quasi-static but
also the isothermal ones, recovering the quasi-static Carnot engine
introduced in section~\ref{sec:quasistatic}, with optimal efficiency $\lim_{(c,d)\to(1,1)} \tilde{\eta} = \eta_{\Carnot}$.
Because of continuity, we can always find values of $c$ and $d$, given a
value of $\chi$, such that the efficiency of our optimal heat engine
is arbitrarily close to the Carnot value and thus greater than the
Curzon-Ahlborn bound. However, it has to be taken into account that
the optimal power for this case is very small, because the denominator
in equation \eref{eq:optimal-power} diverges. In section 1 of the Supplementary Material, we consider the leading order of $\tilde{\eta}$ and $\tilde{\mathcal{P}}$.  

\begin{figure}
  \centering
  \includegraphics[width=\textwidth]{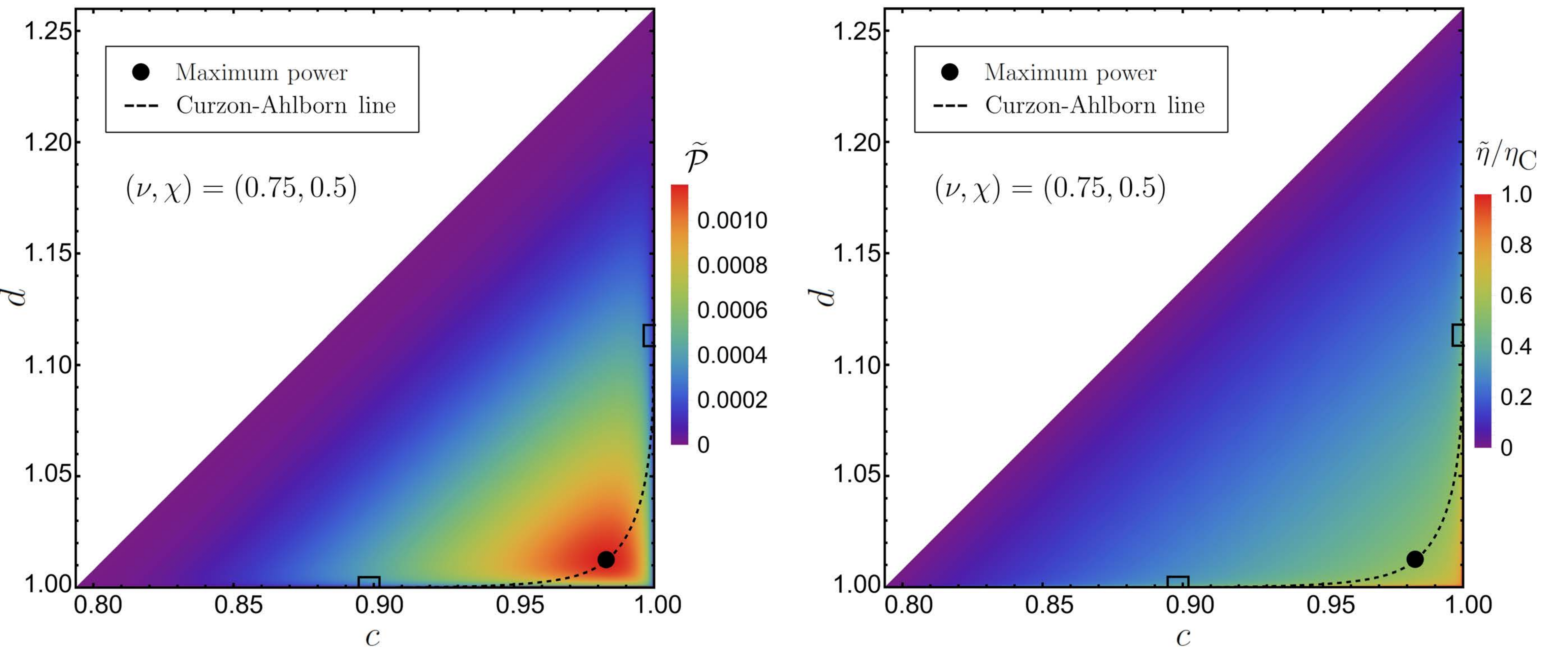}  
  \caption{\label{fig:power-plot-main} Density plots of the optimal
    power (left) and its corresponding efficiency (right) in the
    $(c,d)$ plane.  The curves where $\tilde{\eta}=\eta_{\CA}$ (dashed
    line), with its initial and final points (open squares) over the
    axes $d=1$ and $c=1$, respectively, and the point at which the
    maximum power (circle) is reached, are displayed in both panels. We
    have taken $\nu=0.75$ and $\chi=0.5$. }
\end{figure}

To illustrate the above results, we present in figure
\ref{fig:power-plot-main} a density plot of the optimal power,
equation \eref{eq:optimal-power}, and the corresponding efficiency,
equation \eref{eq:efficiency-at-optimal-power}, as a function of $c$
and $d$. Specifically, we consider given values of the temperature
ratio $\nu=0.75$ and the compression ratio $\chi=0.5$. The constraint
\eref{eq:c-d-chi-constraint} entails that the meaningful region in the
plane $(c,d)$ is a right triangle of vertices
$(c_{\min}=\chi^{\frac{1-\nu}{\nu}},1)$, $(1,1)$ and
$(1,d_{\max}=c_{\min}^{-1})$. Within this region, we can define
another right triangle with the right angle in the same vertex and the
hypotenuse given by the line
$d=\chi^{\frac{\sqrt{\nu}-1}{\sqrt{\nu}}}c$,  above which we know that
$\tilde{\eta}<\eta_{\CA}$. Below the aforementioned line, we cannot
assure that $\tilde{\eta}<\eta_{\CA}$ and in the limit as
$(c,d)\to(1,1)$ we know that $\tilde{\eta}\to\eta_{\Carnot}$.  The
curve over which $\tilde{\eta}=\eta_{\CA}$, which departs from the
hypotenuse vertices  (open squares) of this second triangle  and is fully contained
within it, has been evaluated numerically and plotted (dashed line)
along with the point of delivery of maximum power (circle).

There are several implications that can be drawn from this
analysis. First, along all the sides of the delimiting triangle, the
maximum power is zero because some of the optimal times diverge. 
Second, as a consequence of the previous point and the
positiveness of $\widetilde{\mathcal{P}}$, there always appears a
maximum of the optimal power as a function of $(c,d)$ (for fixed $\nu$
and $\chi$), at a certain point $\tilde{c},\tilde{d}$.  Third, the
numerical estimate for this maximum is very close to the dotted line,
at which $\tilde{\eta}=\eta_{CA}$. This last observation is especially
robust for either small $\chi$ or large $\nu$, as can be seen in
section 2 of the Supplementary Material, in which
analogous plots for different couples of values $(\nu,\chi)$ are
presented.

\subsection{Maximal power for fixed temperature ratio $\nu$}\label{sec:max-power}

The numerical analysis
shown in figure \ref{fig:power-plot-main} suggests
that studying further the maximum power that can be achieved
for fixed values of $\nu$ and $\chi$, that is, as a function of $c$
and $d$, may be illuminating. This is a meaningful physical question: recall that the
reversible Carnot engine is completely determined by these two
parameters. Moreover, its efficiency $\eta_{\Carnot}$ does not depend
on the compression ratio $\chi$, which makes interesting even a further maximisation in the compression ratio $\chi$.

It is possible to address this problem by maximising again the optimal
power in equation \eref{eq:optimal-power} with respect to $c$ and $d$,
and finally with respect to $\chi$. Doing so analytically is not
feasible since it involves transcendental equations. Nevertheless, a
systematic asymptotic analysis can be carried out for $\nu \to 1$. In
this regime, the main idea is to expand all the physical quantities in
powers of $\eta_{\Carnot}=1-\nu$. In order to avoid cluttering the
information flow with the technicalities of the asymptotic
analysis, we present the detailed calculation in sections 3
and 4 of the Supplementary
Material. Therein, it is shown that the expansions of
$\widetilde{\mathcal{P}}$ and $\tilde{\eta}$ in the Carnot efficiency
up to order $\eta_{\Carnot}^{4}$ and $\eta_{\Carnot}^{3}$,
respectively, are
\begin{eqnarray} 
\label{eq:final-optimal-P}
\widetilde{\mathcal{P}}&=&\frac{\eta_{\Carnot}^{2}}{16}- \frac{\eta_{\Carnot}^{5/2}}{8}+\frac{5}{48}\eta_{\Carnot}^{3}-\frac{11}{144}\eta_{\Carnot}^{7/2}+\frac{937}{17280}\eta_{\Carnot}^{4}+O(\eta_{\Carnot}^{9/2}),
\\
\label{eq:final-eta-optimal-P}
\tilde{\eta}&=&\frac{\eta_{\Carnot}}{2}+\frac{\eta_{\Carnot}^{2}}{8}+\frac{\eta_{\Carnot}^{3}}{32}+O(\eta_{\Carnot}^{7/2}).
\end{eqnarray}
We recall that the expansion of the Curzon-Ahlborn efficiency  is
\begin{equation}
\label{eq:Curzon-expansion}
\eta_{\CA}=\frac{\eta_{\Carnot}}{2}+\frac{\eta_{\Carnot}^{2}}{8}+\frac{\eta_{\Carnot}^{3}}{16}+O(\eta_{\Carnot}^{4}),
\end{equation}
Similarly to the situation reported in
Refs.~\cite{schmiedl_efficiency_2008,esposito_thermoelectric_2009} the
first two terms in the expansion of $\tilde{\eta}$ in powers of
$\eta_{\Carnot}$ coincide with those in $\eta_{\CA}$ and the deviation
occurs in the third term, of the order of $O(\eta_{\Carnot}^{3})$. The
obtained efficiency at maximum power is smaller than the
Curzon-Ahlborn bound, similarly to the situation found in
Ref. \cite{schmiedl_efficiency_2008}.\footnote{See equations (24) and
  (25) in that paper. However, the reverse situation has also been
  found, see for instance equation (20) in
  Ref.~\cite{esposito_thermoelectric_2009}.}

\begin{figure}
  \centering
  \includegraphics[width=0.7\textwidth]{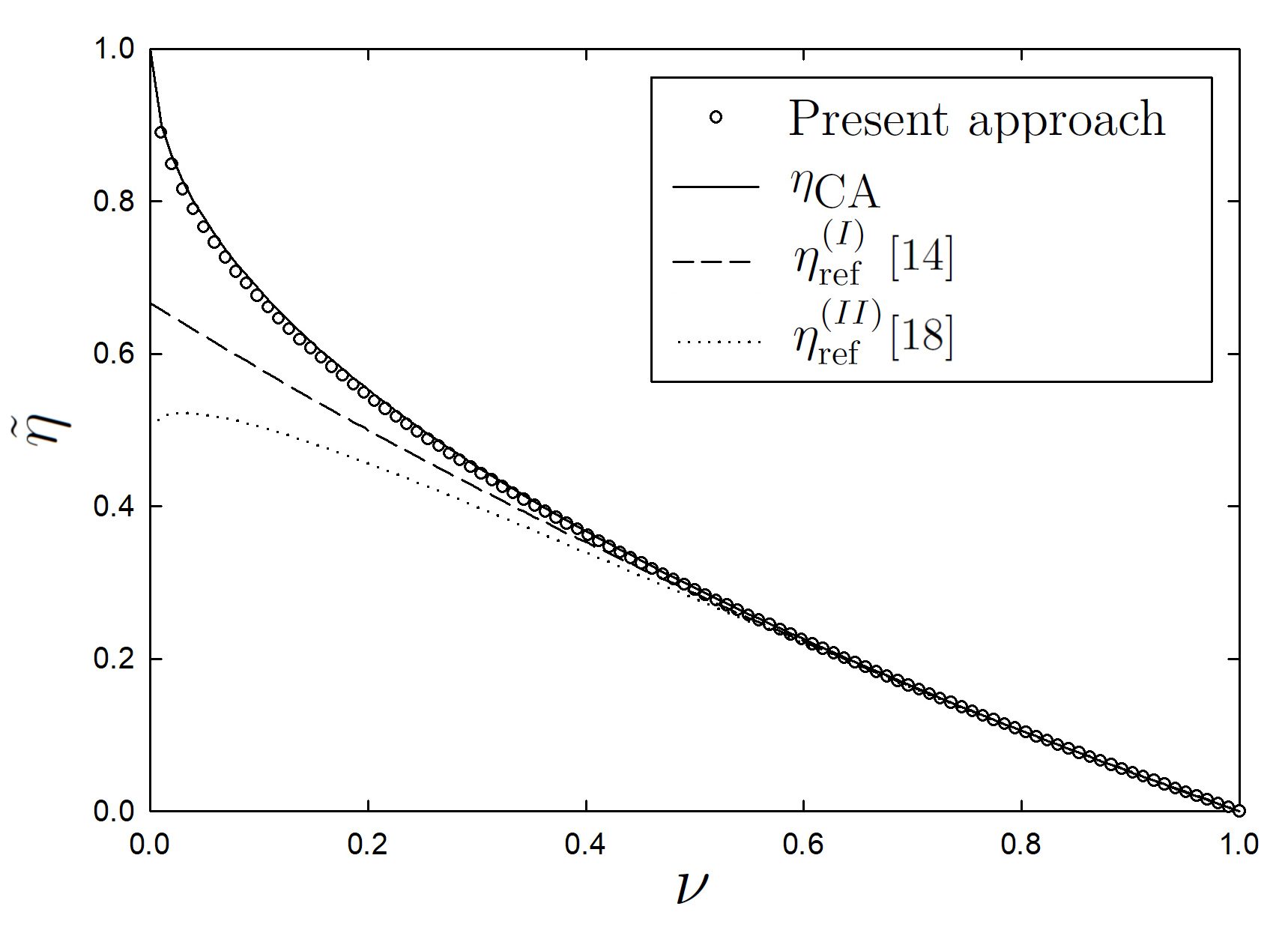}  
  \caption{\label{fig:comparison} Efficiency at maximum power as a
    function of the temperature ratio $\nu$. The value obtained for
    the efficiency, once that the optimisation of the power is
    numerically performed for the rest of parameters, is almost
    indistinguishable from the Curzon-Ahlborn bound $\eta_\CA$. Our
    construction develops a better efficiency compared with those shown
    in Refs.~\cite{schmiedl_efficiency_2008} and \cite{nakamura_fast_2020}. }
\end{figure}

In figure \ref{fig:comparison}, we plot the efficiency at maximum
power as a function of $\nu$. Power has been numerically maximised
over $c,d$ and $\chi$. The obtained efficiency $\tilde{\eta}$ is
compared with (i) the Curzon-Ahlborn bound, (ii) the efficiency for the
engine with instantaneous ``adiabatic'' branches developed in
\cite{schmiedl_efficiency_2008},
$\eta_{\refer}^{(I)}=2\eta_{\Carnot}/(4-\eta_{\Carnot})$, and (iii) the efficiency obtained for large dissipation in the recent proposal,
using a fast forward approach~\cite{nakamura_fast_2020}, 
to build a Carnot-like engine, $\eta_{\refer}^{(II)}= (1-\nu)(1+\sqrt{\nu})/\left[2+\sqrt{\nu}(1+\nu) \right] \leq \eta_{\refer}^{(I)}$.
It is clearly observed that $\tilde{\eta}\geq\eta_{\refer}^{(I)}$ for all
$\nu$, with the difference between them increasing as $\nu$
decreases. Moreover, the closeness between the efficiency of our
engine at maximum power and the Curzon-Ahlborn bound goes beyond our
expectations based on the asymptotic analysis, holding not only within
the limit $\nu \to 1$ but also for the whole range of
$\nu$. Specifically, the relative deviation between our numerical
values for efficiency at maximum power and the Curzon-Ahlborn bound
always remains under  $2\%$.  Therefore, our novel irreversible
Carnot-like heat engine is certainly a very efficient one at maximum
power.

\section{Conclusions}\label{sec:conclusions}

In this work, we have put forward  an irreversible Carnot-like heat
engine. Our model system is a Brownian particle trapped in a harmonic
potential,  in the overdamped regime. The adiabatic branches of the
proposed cycle are \textit{truly adiabatic} in the classical
thermodynamic sense: at every point thereof, there
is no heat exchange with the thermal bath. Of course, the heat
exchange vanishes in average: it is impossible to completely decouple
the colloidal particle from the surrounding fluid. Therefore, our locally
adiabatic branches contrast
with the approach followed in other works, in which the system has a
non-vanishing heat exchange in the ``adiabatic'' parts of the cycle~\cite{schmiedl_efficiency_2008,martinez_adiabatic_2015,martinez_brownian_2016,ciliberto_experiments_2017}. 

The cycle of the reversible Carnot heat engine is completely
characterised by the temperature ratio $\nu$ and the compression ratio
$\chi$. For our irreversible counterpart of the Carnot heat engine, we
need two more parameters in order to fully characterise the four
operating points of the cycle: the adiabatic condition imposes
restrictions on---but does not univocally define---the operating
points.

We have thoroughly studied the performance of the Carnot-like heat
engine at maximum power. We have adopted a step-by-step optimisation
approach. First, the maximum power is shown to be obtained for the
optimal protocols for both isothermal---maximum work
\cite{schmiedl_optimal_2007,schmiedl_efficiency_2008,plata_optimal_2019}---and
adiabatic---minimum duration
\cite{plata_finite-time_2020}---branches. In a second step, we have optimised the
power over the duration of the isothermal processes. These two stages
of the optimisation have been carried out for fixed operation points
in the state space $(\kappa,y,T)$---(stiffness, variance of position,
temperature). Finally, we have maximised the power over the operation
points by just fixing the temperature ratio $\nu$.

The efficiency at maximum power for our heat engine is very close to
the Curzon-Ahlborn bound. This behaviour is predicted by an asymptotic
analysis for $\nu \to 1$. Nevertheless, we have numerically shown that
this result remarkably holds for the whole range of temperature
ratios, well beyond the asymptotic prediction. This implies that our
cycle is a close to optimal choice for building an efficient mesoscopic heat
engine, as compared with the theoretical predictions for other
constructions~\cite{schmiedl_efficiency_2008,nakamura_fast_2020}.
Possible venue for future work includes the study of fluctuations, 
beyond the mean scenario reported here
\cite{gingrich_efficiency_2014,polettini_efficiency_2015}.

\section*{Acknowledgment}

A.P. acknowledges financial support from the Spanish Agencia Estatal de
Investigaci\'on through Grant No.\ PGC2018-093998-B-I00, partially
financed by the European Regional Development Fund. C.A.P. 
acknowledges the support from University of Padova through Project 
No. STARS-Stg (CdA Rep. 40, 23.02.2018) BioReACT grant.
This work has also been financially supported by the
Agence Nationale de la Recherche through Grant No. ANR-18-CE30-0013 (D.G.-O., E.T.).

\section*{References}
\bibliographystyle{iopart-num}

\bibliography{Mi-biblioteca-13-jun-2020}

\end{document}